\documentclass[a4paper,14pt]{article}
\usepackage{soul}
\usepackage[T2A]{fontenc}
\usepackage[english]{babel} 
\usepackage[linktocpage,unicode]{hyperref}
\usepackage{csquotes,graphicx,fancyhdr}
\usepackage{setspace,longtable,graphicx,hyphenat,hyperref,fancyhdr,ifthen,enumitem,amsmath}
\usepackage[dvipsnames]{xcolor}
\usepackage[export]{adjustbox}
\usepackage[linktocpage,unicode]{hyperref}
\hypersetup{unicode=true,colorlinks=true,linkcolor=black,citecolor=blue,urlcolor=blue}
\usepackage{url}
\usepackage{amssymb,amsmath}
\usepackage{geometry}
\usepackage[style=gost-numeric,bibencoding=auto,backend=biber,sorting=none,language=auto,autolang=other,doi=false,url=false]{biblatex}
\AtEveryBibitem{\clearfield{day}
	\clearfield{month}
	\clearfield{endday}
	\clearfield{endmonth}}
\bibliography{Ru-Article_7} 
\usepackage{multido}
\usepackage{appendix}
\newcommand{\e}{\text{e}}
\newcommand{\n}{\text{n}}
\newcommand{\D}{\text{D}}
\newcommand{\M}{\text{M}}
\newcommand{\N}{\text{N}}

\makeatletter
\ams@newcommand{\vardot}[2]{%
  {\mathop{#2\kern0pt}\limits^{\vbox to-1.4\ex@{\kern-\tw@\ex@
   \hbox{\normalfont\multido{}{#1}{.}}\vss}}}}
 \makeatletter
\def\@fnsymbol#1{\ensuremath{\ifcase#1\or \dagger\or  *\or \dagger\dagger
   \or \ddagger\ddagger \else\@ctrerr\fi}}
    \makeatother
\makeatother

\def\nq{\hspace*{-1em}}

\def\beq{\begin{equation}}
\def\eeq{\end{equation}}
\def\lal{&& {}\nq}
\def\bear{\begin{eqnarray}}            
\def\bearr{\begin{eqnarray} \lal}
\def\ear{\end{eqnarray}}               
\def\earn{\nonumber \end{eqnarray}}

\def\det{\mathop{\rm det}\nolimits}

\def\email#1#2{\footnotetext[#1]{e-mail: #2}\addtocounter{footnote}{1}}
\usepackage{color}

\title{Formation of Asymmetrical Two-Brane Structure and its Possible Manifestation}

\author{
Sergey G. Rubin$^{a,b,1}$
        }

\date{\small\it
$^c$  National Research Nuclear University MEPhI (Moscow Engineering Physics Institute),\\ 
            Kashirskoe shosse 31, Moscow 115409, Russia \\
$^d$  N.I. Lobachevsky Institute of Mathematics and Mechanics,
	Kazan  Federal  University, \\
	Kremlyovskaya ulitsa 18,  Kazan 420008,  Russia}

\begin{document}
\maketitle
\email{1}{sergeirubin@list.ru}
\begin{abstract}  
In this paper, we consider the class of extra-dimensional models with two branes and show that each field of the Standard Model must be localized on both neighboring branes, whose asymmetry is of great importance. The discussion is conducted in the framework of a previously developed model. Here we show that the Higgs vacuum average is brane-dependent. As the result, fermion masses on the two branes are also different.
The second brane (brane-2) lacks observers, eliminating the need for fine-tuning; consequently the particle masses remains of the order of the initial energy scale of the universe formation.
Such superheavy charged leptons may serve as a small component of dark matter. Additionally, we show that inter-brane interactions mediated by photons enable massive fermions in brane-2 to act as sources of ultra-high-energy particles. 
The gauge fields are uniformly distributed in the bulk, ensuring charge universality.

\end{abstract}

{keywords: branes, Higgs, fermion, superheavy particles}

 {\begin{itemize}
\item Natural emergence of the 2-brane structure.
\item The Higgs vacuum average is different across branes.
\item Uninhabited brane generates extremely  heavy fermions.
\end{itemize}
}

\section{Introduction}

The concept of supermassive particles—such as neutrinos and charged leptons—is widely used in modern theoretical physics to explain phenomena like the smallness of neutrino masses \cite{Antusch:2005gp}, the Hubble tension \cite{Fernandez-Martinez:2021ypo}, high-energy cosmic rays \cite{Arbuzova:2024uwi}, and dark matter \cite{Song:2023xdk}, among others.
In this article, we explore an extra-dimensional model that inevitably leads to the emergence of supermassive partners of the Standard Model particles.

 The multidimensional gravity is a vital tool for obtaining new theoretical results and explaining known phenomena (\cite{Abbott:1984ba, Brown:2013fba,Chaichian:2000az}). Over the course of several decades, numerous significant results have been achieved, and various concepts have been developed.  The hierarchy problem is undoubtedly one of the most significant issues (see e.g. \cite{Gogberashvili:1998vx,1999PhRvL..83.3370R,ArkaniHamed:1998rs}).  The problem of the small cosmological constant is discussed in a variety of papers, see e.g. \cite{Krause:2000uj, Bronnikov:2023lej}, while the concept of multidimensional inflation is discussed in detail in several sources, including \cite{2002PhRvD..65j5022G,Fabris:2019ecx}. These works assume the extra-dimensional metric is static and stabilization occurs on high-energy scales. The stabilization of extra space as a purely gravitational effect has also been studied in \cite{2003PhRvD..68d4010G,Arbuzov:2021yai}.
Analytical solutions allow for prompt study of the stability \cite{Xu:2014jda,Xu:2022xxd,Liu:2009dt} of one-brane configurations in 5D space-time. 
The stabilization mechanism for the radion in two-brane  models was proposed in \cite{Csaki:1999mp}. The numerical stability of extra dimensions was discussed in our paper \cite{Petriakova:2023klf}.

For several decades, significant interest has been observed in theories concerning the localization of fields on a three-dimensional hypersurface, known as a brane, within a multidimensional spacetime. Akama was the first who introduced this concept \cite{Akama:1982jy}. The hypothesis of the thick brane was independently formulated by Rubakov and Shaposhnikov \cite{Rubakov:1983bb}.

Thin branes, widely discussed in the literature, suffer from some intrinsic problems \cite{Burgess:2015kda}. Therefore, despite hopeful results, they are gradually being replaced by branes with internal structure - thick branes \cite{Bronnikov:2006bu,Chumbes:2011zt,Hashemi_2018,Dzhunushaliev:2019wvv,Bazeia:2022vac,Wan:2020smy}. The latter require special conditions to prove their existence and stability. This could be the choice of a special metric, including the warp factor \cite{Oda_2000} or a special form of the scalar field potential \cite{Bazeia:2022vac}. A purely geometrical approach based on $f(R)$ gravity in 5D is given in \cite{Guo:2023mki}, where a considerable number of bibliographic references can be found.

In brane-world scenarios, the key issue is the localization of fields on branes and the recovery of effective 4D gravity.
The localization of gauge and spinor fields on branes is a necessary element of any brane model and is a widely discussed topic in the literature. The solutions representing the tensor mode excitations in the presence of a brane are examined in \cite{Cui:2020fiz,Xu:2022xxd} with a focus on the stability problem.

The introduction of the warp factor in a D-dim metric \cite{Oda_2000} generalizes the analysis, especially in the fermion sector, as outlined in \cite{Wan:2023usr}, which provides a detailed analysis of this subject in even-dimensional space-time. Fermions in other types of extra metrics are studied in \cite{Gogberashvili_2007,Dantas:2015dca}.

The two-brane ideology is often based on the well-known 5D Randall-Sundrum model (RS1), describing two 3D thin branes located at two fixed points of the orbifold. The model postulates a fixed distance between the branes, which is one of its disadvantages.  A substantial amount of references and ideas concerning this approach can be found in \cite{Olechowski:2024wcf,Tanaka:2000er,Feranie_2010,Dai:2023zsx}. The formation of multiple thick branes is considered in, e.g. \cite{Bronnikov:2007kw,Dzhunushaliev:2019wvv}.
In this study, we base on the results of our paper \cite{Popov:2024nax} where a new class of branes was discussed.
Our analysis there shows that one-brane metrics are exceptions rather than the norm, and we are particularly interested in the set of metrics that describe two-brane configurations.

{Here, we present a model based on our latest research that reveals a two-brane structure. Instead of thin branes introduced as a postulate, this model employs thick branes that arise dynamically from a higher-dimensional $f(R)$ gravity background, with matter confined between them.
}
	
{	Moreover, the present scenario predicts double-peaked localization profiles for all massive Standard Model fields— fermions, and the Higgs field. The model generates an intrinsic asymmetry between brane‑1 and brane‑2, resulting in different physical scales in each brane. This asymmetry leads to rich phenomenological implications, such as the existence of super‑heavy partners of Standard Model particles, and naturally gives rise to the concept of a mirror world discussed in, e.g., \cite{Beradze:2019yyp,Dubrovich:2021gdn}.
}
	 
In this work, we investigate the consequences of a two-brane configuration in six-dimensional spacetime. We demonstrate that:

- {Field Localization and Mass Splitting:} Identical Standard Model fields (neutrinos, charged leptons, and quarks) populate both branes. As energy decreases, each fundamental field undergoes spontaneous splitting into two distinct effective fields with brane-dependent mass spectra.

- {Higgs Mechanism Asymmetry:} The Higgs vacuum expectation value (VEV) differs between the two branes.

- Photons and gravitational waves acting in 4-dim space are distributed uniformly across the extra dimensions (see \cite{Popov:2024nax}), enabling interactions between particles on different branes via photon exchange.

- Inter-brane tunneling of particles is possible but highly suppressed.

- {Gravitational Interaction:} The matter in each brane gravitationally influences the mass distribution of the other brane.

 {A key result of this study is the existence of superheavy partners of Standard Model particles in brane-2. These partners offer a potential explanation for phenomena such as ultra-high-energy cosmic rays and high-energy neutrinos and constitute a subdominant component of dark matter \cite{Khlopov2024,Song:2023xdk}. At present, we cannot constrain their abundance due to the lack of information about their masses and coupling constants.
}


The paper is structured as follows.
In Section 2, we discuss the energy scales suitable for the formation of classical extra-dimensional metrics and the particle content on branes. Section 3 introduces the basis of our model of extra dimensions. Section 4 discusses fermions, photons, and the Higgs field distribution over the extra dimensions. We show that the spinor fields split between the two branes.  It contains the analysis in which cases the matter located on different branes can be described as independent fields and the Higgs mechanism of the mass generation on the both branes. Section 5 discusses the observable signals from the uninhabited brane. Finally, we give a Conclusion in Section 6.

\section{Spreading of matter on branes in the course of their formation}\label{spreadf}


The widely accepted inflationary paradigm implies that the formation of the physical laws takes place above the inflationary scale  {defined by the Hubble parameter} $H_I\sim 10^{14}$GeV. Therefore, action describing field dynamics appears at extremely high energies where the quantum fluctuations fade out. 
The Rundall-Sundrum model with two branes (RS1 model) and its developments \cite{Randall:1999vf} use the second brane as the source of gravity. But there is an aspect which is usually omitted. Indeed, the static two-brane structure does not exist forever, but was formed at high energies. When the universe cools down, two processes occur simultaneously: the extra-dimensional metric describing the branes stabilizes and the fields are randomly settled on them. As a result, the matter fields must be present on both branes. 

 {
	This process is analogous to a second-order phase transition at high temperature. In the non-zero temperature approximation, the effective double-well potential, $V_{\text{eff}}(\Phi, T)$, takes the form
	\begin{equation}
		V_{\text{eff}}(\Phi, T) = V(\Phi) + c T^2 \Phi^2 = \left( c T^2 - \frac{1}{2} \mu^2 \right) \Phi^2 + \frac{1}{4} \lambda \Phi^4
	\end{equation}
Here, $\mu$ is the mass parameter and $\lambda$ the self-coupling constant of the scalar field $\Phi$. In the high-temperature regime $T \gg \mu$, the effective potential is dominated by the thermal term proportional to $T^2 \Phi^2$, with a positive coefficient $c$ that depends on the couplings of $\Phi$ to other fields in the theory.
}
	
 {In the high-temperature regime, the minimum of $V_{\text{eff}}(\Phi, T)$ is located at $\Phi = 0$. As the Universe cools, the coefficient of the $\Phi^2$ term changes sign, leading to the formation of two distinct minima. Consequently, the field rolls into one of these minima, creating domains with different vacuum expectation values (VEVs). The random choice of VEV in a given spatial domain results in neighboring regions with distinct VEVs. These domains are separated by a topological wall, which suppresses interactions between field fluctuations in different domains. Specifically, fluctuations are exponentially attenuated when propagating across the wall. The same physical mechanism underlies the process of brane formation.
}

The two-brane structure is broken by external perturbations if the energy scale is higher than the inverse distance $l_{extra}$ between the two branes. 
So we can consider static metrics at the energies where the wavelengths of the 4D fluctuations are much larger than the extra-space size $l_{extra}$. 
The characteristic wavelengths at the de Sitter stage are of the order of the inverse Hubble parameter $H^{-1}$.  
Therefore, the estimation $H\equiv H_D\ll l_{extra}^{-1}$ provides us with the energy scales where the specific two-brane metric can be considered static.

The D-dim Planck mass $m_D$ determines the highest possible energy scale. So the first limit is $l_{extra}^{-1}\ll m_D$.
Another inequality can be found by using the inflationary concept. It was shown in \cite{Nikulin2019} that a scale of compact extra dimensions cannot be larger than the inflationary scale, $l_{extra}\ll 1/H_I$, in order not to destroy the slow rolling during the inflation. Combining these inequalities, we get the interval for the extra-space scale variation
\begin{equation}\label{l}
H_I\ll l_{extra}^{-1}\ll m_D    
\end{equation}
The whole picture looks as follows. The extra-dimensional metric is formed below the energy scale $m_D\sim 10^{18}$GeV, and is stabilized above the scale $H_I\sim10^{14}$GeV. 
It remains static at lower energies.
The initial fluctuations of the fields are responsible for the stochastic distribution of matter on both branes.

Space-time was created at sub-Planckian energies, as is commonly assumed. This time period is characterized by strong fluctuations of the metric and the fields. The static extra-dimensional metrics, such as the two-brane structure, are formed at much lower energies and are accompanied by fields settling on the branes. The physical parameters characterizing the behavior of a field are contingent upon the properties of the branes.

The hierarchy of the energy scales and the following fine-tuning are necessary phenomena for the formation of complex structures \cite{2007unmu.book..231D}. The selection of the physical parameters for the fine-tuning on a brane is a rather delicate process. It can be done only for one brane and it is reasonable to choose the brane-1 which is populated by observers. brane-2 is then populated by all known sorts of particles with masses much heavier than those on brane-1. As discussed above, the physics as a whole and the brane structure in particular are formed at high energy scale in the interval $10^{14}-10^{18}$GeV. This means that the natural masses of particles on the brane-2 are of the same order of magnitude. 

We come to conclusion that there are all sorts of known particles on the brane-2 which are many order of magnitude more massive compared with those on the brane-1. The matter on the brane-2 is incapable of forming complex structures because of the lack of the fine-tuning.

The potential contribution of such massive particles to the dark matter sector warrants investigation. The following subsection presents a general analysis of this hypothesis, decoupled from specific brane-world configurations. We assume that the gravitational and electromagnetic forces between matter are invariant with respect to the brane on which the matter is located. This assertion, was rigorously established in \cite{Popov:2024nax}.

\section{Two-brane metric. }

\subsection{The model choice}

The presence of at least two branes is necessary element of this study so that we will base on the model elaborated in \cite{Popov:2024nax}. In that study, we impose the extra-dimensional  metric and show that a two-brane structure is formed under some assumptions which are not very restrictive.
The metric is supposed in the form
\begin{equation}\label{metric}
ds^2=g_{AB}dX^AdX^B=\e^{2 \gamma(u)}\left[dt^2 -\e^{2Ht}(dx^2 +dy^2 +dz^2)\right] - du^2 -r(u)^2 d \Omega_{n-1}^2. 
\end{equation}
 {Here we consider a 6D space so that $D=4+n,\, n=2$ and the set of extra-dimensional coordinates is $y=\{X^4,X^5\}\equiv \{u,\theta\}$.
$d \Omega_{n-1}^2$ is the metric on a unit $n-1$-dimensional sphere. The angular coordinate $X^5=\theta$ varies in the finite region so that the solid angle $d\Omega_{n-1}=d\theta$}; $X^{\mu}\equiv x^{\mu}, \mu=0,1,2,3$.
We do not fix the limits of the radial coordinate $u$ a priori.
The Hubble parameter $H$ is arbitrary.

Consider $f(R)$ gravity in a $\D = 4 + n$-dimensional manifold $M_\D$: 
\begin{eqnarray}\label{SfR0}
S = \frac{m_{\D}^{\D-2}}{2}  \int_{M_\D}  d^{\D} X \sqrt{|g_{\D}|}  f(R) ,
 \quad f(R)=aR^2+R+c,
\end{eqnarray}
where $g_{\D} \equiv \det g_{\M\N}$, $\M,\N =\overline{1,\D}$, 
the $n-1$-dimensional manifold $M_n$ is assumed to be closed, $f(R)$ is a function of the 
D-dimensional Ricci scalar $R$, and $m_\D$ is the $\D$-dimensional Planck mass.

Remark that there are still no traces of branes, neither in the metric \eqref{metric} nor in the gravity action \eqref{SfR0}. Nevertheless, branes are inevitably formed during the cooling of the universe. See \cite{Popov:2024nax}.
The two-brane structure reveals as solutions of the classical equations

\begin{eqnarray}  \label{eqMgravity}
    -\frac{1}{2}{f}(R)\delta_{\N}^{\M} + \Bigl(R_{\N}^{\M} +\nabla^{\M}\nabla_{\N} 
        - \delta_{\N}^{\M} \Box_{\D} \Bigr) {f}_R  =0
\end{eqnarray}
  with $f_R = {df(R)}/{dR}$, $\Box_{\D}= \nabla^{\M} \nabla_{\M}$. 

\begin{figure}
    \centering
\includegraphics[width=0.45\linewidth]{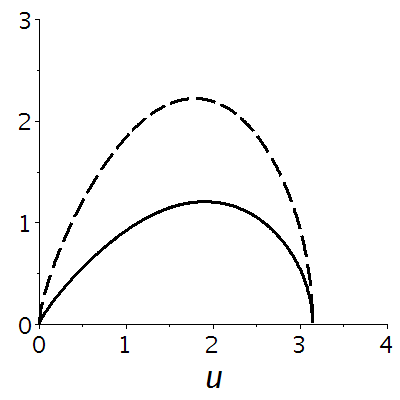}
    \caption{The typical metric functions, $\gamma(u)$ - solid line, $r(u)$ - dashed line. The interval is normalized so that the endpoints are $u_1=0,\quad u_2=\pi$ in the units $m_D=1$. Their shapes depends on additional conditions and are characterized by zeros at the endpoints. The parameter values are $p= 1/3 , q=1/5 , A=2$.   {As was shown in \cite{Popov:2024nax}, test particles move toward the endpoints ($u=0$ or $u=\pi$ in this case) along the geodesics derived from the metric containing function $\gamma(u)$.}}
    \label{metricfig}
\end{figure}

The formation of the two-brane static structure based on $f(R)$ gravity, as in Eq. \ref{eqMgravity}, with the parameters $a=300$ and $c=0.002$, was discussed in our previous papers \cite{Rubin:2015pqa,Nikulin2019} where additional (boundary) conditions necessary to find a particular solution were also considered. In fact, solutions containing two branes have been found for a wide range of parameters and additional conditions.
It was shown there that their common form is characterized by two zeros of the metric functions $r(u_{1,2}), \gamma(u_{1,2})$, which is illustrated in Figs.\ref{metricfig}. The manifold is a compact n-dimensional space of a finite volume. 
Here we use a parametric set of metrics with two brane structure 
\begin{eqnarray}\label{metparam}
    && e^{\gamma(u)}=u^p\cdot \sqrt{\sin(u)},\nonumber \\
    &&r(u)=Au^q\cdot \sqrt{\sin(u)}.
\end{eqnarray}
They retain the above properties being more clear for the analysis. The distance between the branes is the matter of calculations and depends on the additional conditions. We renormalize this distance to the standard interval $(0, \pi)$.

The classical equations are senseless at the distances $m_D^{-1}$ where the quantum regime dominates. This statement holds for any classical system and is the reason why the space-time quantum foam exists at the Planck energy scale. In our case, this means that a brane width should be substantially larger than $m_D^{-1}$.

The thickness of the brane is related to the fields localization in its vicinity. As was shown in \cite{Bronnikov:2023lej, Popov:2024nax}, the classical field distributions along the extra coordinate $u$ are characterized by two areas of localization.
One can conclude that the model described above leads to the formation of the two-brane structure. 

\subsection{The Planck mass}

In this paper, we widely use the units $m_D=1$ and it is instructive to restore its dimensionality at the final stage.
The $D$-dimensional Planck mass $m_D$ can be fixed by its connection to the known four-dimensional Planck mass $m_4$.
For this purpose, we make the necessary preparations here, starting with the representation of the Ricci scalar of 6D extra space in the form
\begin{eqnarray}
    &&R\equiv R_6=e^{-2\gamma(u)}R_4(x)+R_2(u)+\mathcal{R}(u), \\
    &&\mathcal{R}=-\frac{8}{r(u)}\gamma' (u)r'(u),\\
    &&R_2(u) = -\frac{2r(u)''}{r(u)},
\end{eqnarray}
where 2D extra metric presented in \eqref{metric} has been taken into account. Here $R_4$ is the Ricci scalar of our 4D space and $R_2$ is the Ricci scalar of the extra space.  We also assume that the distributions of matter and metric functions over the extra dimensions depend only on the radial extra coordinate $u$. In this case, the action \eqref{SfR0} takes the form
\begin{eqnarray}\label{SfR}
&&S = \frac{m_{\D}^{\D-2}}{2}  \int_{M_\D}  d^{\D} X \sqrt{|g_{\D}|} \,   f(e^{-2\gamma(u)}R_4(x)+R_2(u)+\mathcal{R}(u)) \simeq \nonumber\\
&&\frac{m_{\D}^{\D-2}}{2}  \int_{M_\D}  d^{4} x d^n y \sqrt{|g_{\D}|} \, [f(R_2(u)+\mathcal{R}(u)) + e^{-2\gamma(u)}f_R(R_2(u)+\mathcal{R}(u))R_4(x)+ O(R_4^2)] \simeq \nonumber\\
&& \frac{m_4^2}{2}\int d^4x \sqrt{|g_{4}|} [R_4(x)-2\Lambda]
\end{eqnarray}
The 4D Planck mass $m_4$ is related to the D-dim Planck mass $m_D$ as follows:
\begin{equation}\label{m4D}
m_4^2=m_D^{D-2}v_{n-1}\int du e^{2\gamma(u)}r(u)f_R(R_2+\mathcal{R}),
\end{equation}
where $v_{n-1}$ is the volume of an $n-1$-dim sphere of unit radius 
\begin{equation}\label{vn-1}
v_{n-1} = \frac{2\pi^{\frac{n}{2}}}{\Gamma(\frac{n}{2})}.    
\end{equation}
The Planck mass $m_4(H=0)=m_{Pl}$ is measured at the low energies, where we assume that the inequality $R_4\ll R_2$ holds.

Let us estimate the value $m_D$. To this end, suppose that all dimensional values are of the order of $m^D$, so that
\begin{equation}\label{mD4approx}
m_D\sim m_4/\sqrt{v_{n-1}}.   
\end{equation}
The volume \eqref{vn-1} has maximum $v_{n-1}\simeq 33$ at $n\simeq 7$ which gives the estimation of the smallest value of $m_D\sim 0.2m_4\simeq 10^{18}$GeV. Here and below we suppose $m_D\sim 10^{17}\div 10^{18}$GeV and the scale of the two-brane structure $l_{extra}\sim  10^{-14}\div 10^{-15}$GeV$^{-1}$ for the estimations.
Other terms in \eqref{SfR} contribute to the vacuum energy density $\Lambda$, see discussion in \cite{Bronnikov:2023lej} , which is not used here. 
The knowledge of the metric functions $\gamma(u), r(u)$ as in Figure \ref{metricfig} permits to restore the 4D Planck mass $m_D$, see expression \eqref{m4D}. 

A non-trivial scenario arises if the number $n$ of extra dimensions is large. In this case, the volume of extra dimensions \eqref{vn-1} is a rapidly decreasing function of the number of extra dimensions. For example, $v_{n}\simeq 6.5\cdot 10^{-4}$ at $n=3$, so
$$m_D\sim 10^2 m_4\sim 10^{21}\text{GeV}.$$
The onset of the quantum regime occurs at a scale $ 1/m_D \sim 10^{-2}/m_4$, which is significantly smaller than the typically mentioned estimation $1/m_4$.




\section{Fields splitting between two branes}\label{split}

Here we study the distributions of fields over the two branes and their features leading to observational consequences after the reduction to the 4D physics. We also discuss conditions under which a specific field disposed on two branes can be considered as two fields with effectively different parameters.

The reduction to 4 dimensions assumes integration over the extra coordinates. So we have to know the distributions of fields (scalars, fermions, vectors) within the extra dimensions. The suitable way of a field decomposition is as follows:
\begin{equation}\label{decomp}
\Phi(x,y)=\Upsilon(x)Y_{cl}(y)+\sum_{k=0}^{\infty} \Upsilon_k(x)Y_k(y)\simeq \Upsilon(x)Y_{cl}(y),
\end{equation}
where $\{Y_k(y)\}$ is a set of orthonormalized functions  {describing the extra metric excitations which are neglected if the energy is not extremely high. The normalization constant is hidden in the field $\Upsilon(x)$ and is used at the final state to convert the 4D kinetic term to its standard form,  for example  $\frac 12 \partial_{\mu}\Upsilon(x)\partial^{\mu}\Upsilon(x)$, in case of the scalar field.}
It is usually assumed that they satisfy the equation
\begin{equation}\label{box}
    \Box_n Y_k(y) = \lambda_k Y_k(y), k=0,1,2, ...
\end{equation}
The zero mode $Y_0(y)=1$ satisfies this equation with $\lambda_0 = 0$.
The function $Y_{cl}(y)$ is the classical solution to the field equation(s) provided that the derivatives in $x$ coordinates of our space are neglected. This approximation holds if the wavelengths of the perturbations acting in our 3-dim space are much larger than a scale of the extra dimensions. 
It also means that the sum describing the field fluctuations along the extra dimensions can be omitted. The function $\Upsilon$ is assumed to be endowed with group indices depending on the group representation of the field while the scalar function $Y_{cl}$ is responsible for the distribution over the radial extra dimension.  The fields are treated as trial functions in this research. This step does not involve any fine-tuning. The amplitude of the classical distributions, $Y_{cl} (u)$, is of the order of the D-dimensional Planck scale, $m_D$.

\subsection{Fermions distribution over the extra dimensions.}\label{fermi}

In this subsection we shortly remind the results of \cite{Popov:2024nax} concerning the distribution of fermions along the radial extra-dimensional coordinate $u$ near the brane and in the bulk. General form of the free massless fermion action is 
\begin{equation}\label{Sff}
S_{fermion}=\int d^D z\sqrt{|g_D|}\bar{\Psi}i\Gamma^M D_M \Psi 
\end{equation}
The mass term is assumed to be generated by the interaction with the Higgs field.
The particular form of the curved gamma matrices $\Gamma^M$ and the covariant derivatives can be found in \cite{Oda_2000}. The useful discussion is also given in \cite{Wan:2023usr,Appelquist:2000nn,Randjbar-Daemi:2000lem}.
 
\begin{figure}
    \centering
    \includegraphics[width=0.5\linewidth]{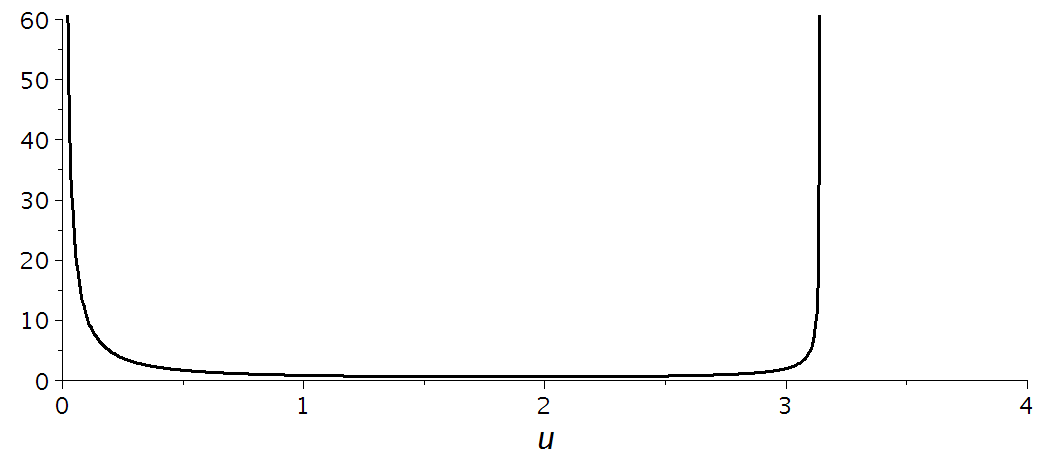}
    \caption{Fermion distribution over the extra dimensions. The metric functions are represented in Fig.\ref{metricfig}. The localizations near $u_1=0$ and $u_2=\pi$ are evident.}
    \label{fermifig}
\end{figure}

 {As shown in \cite{Gogberashvili_2007}, a 6D spinor in the s-wave approximation 
\begin{equation}\label{2bispinor}
    \Psi =\left(\begin{array}{c}
  \xi_1(x^\mu) \\
  \xi_2 (x^\mu)
\end{array}\right)Y_f(u) \equiv \psi(x)Y_f(u)
\end{equation}
is equivalent to a pair of 4D Dirac spinors, $\xi_1$ and $\xi_2$, which are equal in this case. Here the  normalized function  $\psi(x)$ is endowed with spinor and group indices acting in 4D space. 
This study focuses on the distribution of fields  $Y_f(u)$ across the extra-dimensional coordinates rather than their reduction to four-dimensional physics. 
Consequently, our attention is directed towards the function $Y_f(u)$ in the form
\begin{equation}\label{Yfermi}
	Y_f(u)=r(u)^{-1/2}e^{-2\gamma(u)},
\end{equation}
see Fig.\ref{fermifig},  which is a solution to equation
\begin{equation} \label{eqY}\Big[\partial_u + \frac{r'(u)}{2r(u)} + 2 \gamma'(u)\Big]Y_f(u)=0.\end{equation}
This equation was thoroughly derived in \cite{Oda_2000}.
}
%
The particular distribution over the radial coordinate $u$ is shown in Fig.\ref{fermifig}.  { This distribution has two sharp peaks at the coordinates $u_1$ and $u_2$ as can be seen also from the analytical expression \eqref{Yfermi} and Fig. \ref{metricfig}. 
}
The region proximate to $u_1$ is designated as brane 1, and the area near $u_2$ is labeled as brane 2. 
The fields on both branes act independently because the distribution between them is almost zero. 
A rigorous {proof} of this statement is in the next subsection. 
The field behavior on brane~1 and brane~2 is described by the distribution in Fig.~\ref{fermifig}, 
spanning the intervals $(0, \sim 1)$ and $(\sim 2, \pi)$, respectively.


%

An important fact is that the observable fermions are described by the wave function $\psi(x)$ in the decomposition $\Psi(x,u) = \psi(x) Y_f(u)$ and propagate in the four-dimensional spacetime orthogonal to the extra coordinate $u$. The function $Y_f(u)$ exhibits two distinct peaks near the endpoints $u_{1,2}$.
Thus, the fermion field density is proportional to $|Y_f(u)|^2$ and is  maximal near the endpoints $u_{1,2}$.

Also, $Y_f(u) \to 0$ (or is approximately vanishing) in the interbrane region, so that the fermion field density tends to zero there. As a result, the fermion excitations $\psi(x)$ on each brane do not influence each other and can be treated as independent.
 A rigorous proof follows in the next subsection.


\subsection{Fields distribution in branes}\label{spread}

Consider an action featuring the specific field $\zeta(x,y)$ along with its associated set of Lagrangian parameters $\{\lambda_{in}\}$, which include mass and coupling constants. Let us find the conditions for which this field is observed at low energy as two independent fields characterized by different sets of parameters $\{\lambda_{in}\}\to\{\lambda_{1,2}\}$. The transition amplitude is calculated using the path integral approach,
\begin{equation}\label{Ampl0}
    A=N\int D\zeta (x,y)\exp{\left(iS[\zeta;\lambda_{in}]\right)}, \quad D\zeta (x,y)=\prod_{x,y}d\zeta(x,y) .
\end{equation}
Here $\lambda_{in}$ is an initial set of physical parameters. An action $S$ in \eqref{Ampl0} is a functional of a function $\zeta$ acting in D-dim.
At the highest energy levels, where $H\sim m_D$, this field is dispersed throughout the entire extra space. However, as the system's energy decreases to lower levels, the scenario changes. As was discussed in the section \ref{spreadf}, the extra space structure is formed at the scale $l_e\lesssim 10^{-27}$cm. At the present time, the smallest scale where achieved at the colliders $l\sim 10^{18}$cm. The inequality $l_e\ll l_s$ allows to neglect the 4D variation of fields and use the classical equations for fixing the extra-dimensional metric and the fields distribution over extra coordinates $y$. The decomposition  \eqref{decomp} in the form
\begin{equation}\label{decomp3}
\zeta(x,y)=\Upsilon (x)Y(y)
\end{equation}
appears to be useful here. So we can apply the  stationary phase method and substitute the distribution $Y(y)$ by the classical one $Y(y)\to Y_{cl}(y)$ over the extra dimensions and leaving the 4D distributions $\Upsilon(x)$ arbitrary. The example of the classical equation admitting analytical solution due to its simplicity, is written in the subsection \ref{fermi}. 
The amplitude has the form
\begin{eqnarray}\label{Ampl1}
&& A=N\int D\zeta (x,y)\exp{\left(iS[\zeta;\lambda_{in}]\right)}= N\int \prod_{x}d\Upsilon(x)\prod_{y}dY(y)   \exp{\left(iS[\Upsilon(x)Y(y);\lambda_{in}]\right)}\simeq \nonumber\\
&&N'\int \prod_{x}d\Upsilon(x)  \exp{\left(iS[\Upsilon(x)Y_{cl}(y);\lambda_{in}]\right)}=N'\int D\Upsilon \exp{\left(iS^{(4)}[\Upsilon(x));\lambda_{eff}]\right)}
,
\end{eqnarray}
where the main order of the approximation is used in the last line.
Knowledge of the function $Y_{cl}(y)$ permits to integrate over the extra dimensional coordinate $y$ which lead to the renormalization of the initial physical parameters values $\lambda_{in}\to\lambda_{eff}$ in effective 4D action $S^{(4)}$.  The general conclusion is that the low energy effective parameters $\lambda_{eff}$ appears to be dependent on accidental field distribution at high energies. 
This result being very important, has been performed for various fields in \cite{Rubin:2016ude,Petriakova:2023klf, Bronnikov:2023lej,Popov:2024nax}.

In the papers mentioned above, it was also shown that the spinor, scalar and the Higgs field distributions has sharp peak near end points, while the intermediate region is characterized by extremely small field value. Fig.\ref{fermifig} illustrates the spinor field localization near each endpoint. This remark opens new opportunity in the research and we consider it more thoroughly below.
The region between end points $(u_1, u_2)$ can be divided into three distinct parts. The first region $(u_1, u_1')$, denoted as $\mathcal{U}_1$, encompasses the extra-dimensional radial coordinate $u \in \mathcal{U}_1$ forming the brane-1, while another region $(u_2',u_2)$ includes the coordinate $u \in \mathcal{U}_2$ and forms brane-2. Both of this regions are characterized by sharp peak (casp) in the field distribution.
The third region $(u_1', u_2')$, denoted by $\mathcal{U}_0$, acts as a separator between the two branes. Here, the field and its derivatives are negligible, and hence this part of the field action $S_0 \simeq 0$ with appropriate accuracy. The precise definition of the boundaries between the regions $\mathcal{U}_1$, $\mathcal{U}_0$, and $\mathcal{U}_2$ is not necessary. 

\subsection{Splitting a field into two independent effective fields}
 {Here, we demonstrate that a field distributed across both branes can effectively be treated as two independent fields.}
The first step in application of the Method of Stationary Phase is to find classical function delivering an extrema of action. Assume that this function has two sharp maxima $\mathcal{U}_1$ and $\mathcal{U}_2$ and is (almost) zero between them region $\mathcal{U}_0$, like in Fig.\ref{fermifig}.
Then, the action can be segmented into three components based on integration over the regions $\mathcal{U}_1$, $\mathcal{U}_0$, and $\mathcal{U}_2$
Taking all this into consideration, the transition amplitude \eqref{Ampl0}
\begin{eqnarray}\label{Ampl12}
A=N\int D\zeta_1 \exp{(iS_1[\zeta_1]})\int D\zeta_0  \exp{(iS_0[\zeta_0]}) \int D\zeta_2   \exp{(iS_2[\zeta_2]})
\end{eqnarray}
\begin{eqnarray}
&&D\zeta_k=\prod_{x,y,u\in \mathcal{U}_k}d\zeta_k(x,y), \quad k=0,1,2.
\end{eqnarray}
 Measure $D\zeta_0$ includes the separating points $\zeta_1'$ and $\zeta_2'$. Now we can apply the Method of Stationary Phase for three integral in the same manner as it has been done in \eqref{Ampl1} to obtain
 \begin{equation}\label{Ampl2}
     A=N'\int \prod_{x}d\Upsilon_1(x)  \exp{\left(iS[\Upsilon_1(x)Y_{1,cl}(y);\lambda_{in}]\right)}\int \prod_{x}d\Upsilon_2(x)  \exp{\left(iS[\Upsilon_2(x)Y_{2,cl}(y);\lambda_{in}]\right)}
 \end{equation}
 The function $\Upsilon(x)$ is denoted as $\Upsilon_1(x)$ in the region $\mathcal{U}_1$ and similarly for other regions. The second of the three integrals, $S_0$, is essentially zero owing to the negligible field value $Y_{cl}(y)$ and its derivatives between the branes (region $\mathcal{U}_0$).
Knowledge of the classical functions $Y_{1,2,cl}(y)$ allows one to perform the integration of the actions $S_1$ and $S_2$ over the extra coordinates $y$ which leads to the amplitude
 \begin{equation}\label{Ampl3}
     A=N'\int \prod_{x}d\Upsilon_1(x)  \exp{\left(iS^{(4)}[\Upsilon_1(x);\lambda_{eff,1}]\right)}\int \prod_{x}d\Upsilon_2(x)  \exp{\left(iS^{(4)}[\Upsilon_2(x);\lambda_{eff,2}]\right)}
 \end{equation}
for the fields $\zeta_1, \zeta_2$ acting in the disconnected areas $\mathcal{U}_1$ and $\mathcal{U}_2$. The normalization constant is changed as
$$N'=N\int D\zeta_0 .$$
The integration here is performed over the spatial domain $\mathcal{U}_0$, where $S_0=0$.

Formally, transition amplitude \eqref{Ampl3} describes two independent fields $\Upsilon_1$ and $\Upsilon_2$ acting in 4-dim. 
An essential fact is the difference in the parameter values of the fields, $\lambda_{eff,1}\neq \lambda_{eff,2}$, located near different branes. Indeed, the integration over the extra dimensional coordinates $y$ leads to a transformation of the initial parameter values $\lambda_{in}$ in \eqref{Ampl2} into the effective parameters $\lambda_{eff,1}$ for the fields on brane-1 and $\lambda_{eff,2}$ for brane-2.

the two localized fields interact negligibly due to exponentially suppressed overlap. However, there are waves that could transfer energy between the branes. At the low energies, the wavelengths of external excitations are much larger compared to the extra space scale and hence cannot be excited.

\subsection{Weak interaction of particles localized in different branes}\label{floc}

The existence of a region characterized by an extremely small amplitude of a field between the branes as in Fig.\ref{fermifig}, leads the bifurcation of the field into two effective fields with distinct parameters, which has profound implications.  
This subject is examined in detail in the previous subsection. It is shown therein that a field situated on distinct branes can be treated as two independent fields, each with its own set of the physical parameters. 

The explanation on the ''physical level'' attracts the analogy with a scalar field behavior within the double-well asymmetric potential. 
%
When the energy of the system tends to zero, the field is concentrated near the potential minimums. In this case, the small-amplitude fluctuations around each of the minima are observed as free scalar particles with specific masses depending on the shapes of the minima. 

A maximum of potential is responsible for the exponential attenuation of the field amplitude between the two minima. The overlapping integral of the fluctuations near distinct minima is small, so that the fluctuations around both minima do not interact with each other and can be considered as independent fields.

The same is true for fields concentrated near branes. The field near one brane interacts extremely weakly with the field in the other brane due to the smallness of the overlap integral. This is true for the zero mode $y_{cl}$. The  KK mode excitations are forbidden if their typical wavelength  $l\gg l_{extra}$,\, ($l_{extra}$ is the distance between the branes) is much smaller than the electroweak scale.

The field distributions near the two branes can be independently approximated by an appropriate sets of orthonormalized functions specific for each brane. These could be the sets built on the basis of harmonic oscillator, or the Coulomb set of wave functions for the hydrogen atom, for example. 
The general form of the wave function for the n-th energy level of a harmonic oscillator is as follows
\begin{equation}\label{set}
\phi_n(u) = N_n H_n(\alpha u) e^{\alpha^2u^2/2}  
\end{equation}
where $N_n$ is the normalization factor, $H_n$ is the Hermite polynomial of the n-th order, $\alpha$ is an arbitrary parameter.
In the following, we will assume that all distributions are described by the set of functions $\phi_n(u-u_1)$ if they are located near brane-1, and functions $\phi_n(u_2-u)$ if they are located near brane-2. 

It is assumed that the observers are located at the brane-1. In this case, it is appropriate to choose the field distribution as a sum over the orthonormal basis \eqref{set}
\begin{equation}\label{Y1}
   Y_f^{(1)}(u)=\sum_{n} a_n\phi_n (u-u_1) 
\end{equation}
which reflects the fact of the field localization around $u=u_1$ (brane-1). 
A similar expression can be written for the fermion field located on brane-2 at $u=u_2$,
\begin{equation}\label{Y2}
    Y_f^{(2)}(u)=\sum_{n} b_n\phi_n (u_2-u) .
\end{equation}
These functions quickly tend to zero if we limit ourselves by several terms in the sums. Thus, the distribution over the whole interval $(u_1, u_2)$ can be approximated with good accuracy in the form
\begin{equation}\label{Psi12}
    \Psi(x,u)=\Psi^{(1)}(x,u)+\Psi^{(2)}(x,u)=\psi_1 (x)Y_f^{(1)}(u)+\psi_2 (x)Y_f^{(2)}(u)
\end{equation}
The first term has a sharp maximum at the brane-1, while the second term has a sharp maximum at the brane-2.
This approximation is valid for all sorts of fields, provided they are localized on the different branes. Substitution \eqref{Psi12} into the action \eqref{Sff} leads to the effective action for the two free fermions $\Psi^{(1,2)}$. For example, the interference term $\bar{\Psi^{(1)}}iD_A\Gamma^A\Psi^{(2)}$ is proportional to the overlapping integral
\begin{equation}\label{overl}
 \int du \sqrt{g_n} Y_f^{(1)}(u)Y_f^{(2)}(u)
\end{equation}
an hence is negligibly small. The main result of this section is the following: the fermions concentrated around different branes do not influence each other and can be considered as two different sorts of fermions.

\subsection{Photon interaction}

In this subsection, we use the results obtained in \cite{Cui:2020fiz, Popov:2024nax} where it was shown that the electromagnetic field and the gravitational excitations are distributed uniformly between the branes. 


Here we address two questions: (1) Is the electric charge of electrons identical across different branes? (2) Can electrons transit between branes?
So, we have to consider a D-dim extension of the SM action for the electron-photon interaction
\begin{equation}\label{Sint}
S_{int}\equiv\alpha_D {v}_{\n-1}\int \sqrt{|g_4|}d^4x \int_{u_1}^{u_2}  d u\sqrt{|g_n|}\bar{\Psi}\Gamma^N A_N \Psi .
\end{equation}
Here $\alpha_D$ is a coupling constant in D dimensions and $A_N$ is the gauge field. It is shown in \cite{Popov:2024nax} that the gauge field is uniformly distributed over the extra dimensions, i.e. $A_N(x,y)=Const\cdot a_N(x)$. 
Action \eqref{Sint} can be reduced to the 4-dim action:
\begin{eqnarray}\label{fe0}
S_{int}\simeq\alpha_D {v}_{\n-1}  \int \sqrt{|g_4|}d^4x \int_{u_1}^{u_2}  du\sqrt{|g_n|}\left(\bar{\Psi}_1\Gamma^N A_N \Psi_1  + \bar{\Psi}_2\Gamma^N A_N \Psi_2 + \bar{\Psi}_1\Gamma^N A_N \Psi_2 + h.c.\right)
\end{eqnarray}
where $\Psi=\Psi_{1}+\Psi_{2}$ and the functions $\Psi_{1}$, $\Psi_{2}$ are located on the brane-1 and brane-2 correspondingly, according to discussion in subsection 4.2. The next step is to substitute the decompositions $\Psi_{1,2}=\psi_{1,2}(x)\cdot Y_f^{(1,2)}(u)$ into action \eqref{fe0}, to obtain
\begin{eqnarray}\label{fe}
S_{int} = \int \sqrt{|g_4|}d^4x \left(\alpha\bar{\psi}_1\gamma^{\mu} a_{\mu} \psi_1  +  \alpha \bar{\psi}_2\gamma^{\mu} a_{\mu} \psi_2 + \alpha_{12} \bar{\psi}_1\gamma^{\mu} a_{\mu} \psi_2 + h.c.\right)
\end{eqnarray}
where expression \eqref{Psi12} have been kept in mind. 
We also used the fact that the extra space distribution of the fermions is included into the normalization of the wave functions $\psi_{1,2}$ and $\alpha=Const^2 \alpha_D$. The observable coupling constant $ \alpha$ is the same on both branes. The parameter $\alpha_{12}$ is proportional to the overlapping integral \eqref{overl} and  hence, is extremely small.

Action \eqref{fe} contains  the vertices describing the interaction of photons with ''normal'' electrons $\psi_1$ located at the brane-1 (first term) and ''heavy'' electrons $\psi_2$ on the brane-2 (second term).  A photon emitted from the brane-1 can be absorbed by fermions located on brane-2, and vice versa. We will use this fact below. 

The third term represents the vertex containing both types of electrons in conjunction with the electromagnetic field. Hence, the transition of the heavy electron to another brane, accompanied by photon emission, is associated with this vertex. This effect is negligibly small due to the smallness of the coupling constant $\alpha_{12}$.


\subsection{The Higgs mechanism in different branes}\label{Higgs}

As was mentioned in subsection \ref{spread}, the brane other than that settled by observers are most likely filed by heavy partners of the SM particles. According to the Standard Model, the Higgs field is responsible for the fermion masses. Therefore, it is of interest to study the extra dimensional distribution of the Higgs field and its interaction with the fermions. 

\begin{figure}
    \centering
    \includegraphics[width=0.8\linewidth]{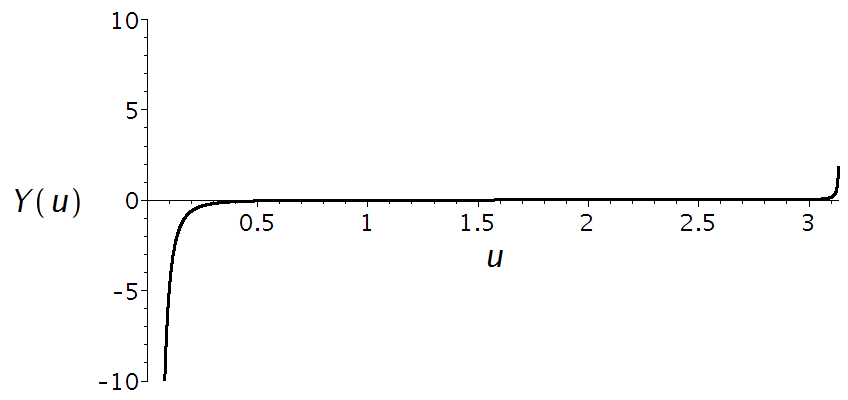}
    \caption{The Higgs field distribution over the internal coordinate $u$ is specific to each brane. The background metric functions are represented in Fig.\ref{metricfig}. Additional conditions are $Y_H(1.5)=0,\, Y_H'(1.5)=10^{-2}$.  {The sharp peaks which define the branes are the result of numerical solutions to equation \eqref{boxUU} in wide range of parameters. The Higgs field is approximately zero  in the interval between two black points. The fermion fields are also concentrated near the branes, see Fig.\ref{fermifig}. }
     }
    \label{Higgsfig}
\end{figure}

The distribution of the Higgs field 
\begin{equation}\label{Hvro}
    H(x,y)=h(x)Y_H(u),\quad h(x) = \frac{1}{\sqrt{2}} 
\begin{pmatrix}
0 \\ v_h +\rho(x)
\end{pmatrix}, \quad v_h \gg\rho(x)
\end{equation} 
in the form \eqref{decomp} over the extra coordinate $u$ is governed by the scalar function $Y_H(u)$ which is the solution to the classical equation
\begin{equation}        \label{boxUU}
-\left[\partial^2_u +\left(4\gamma'+\dfrac{r'}{r}\right)\partial_u\right] Y_H(u) =\nu Y_H(u) - \lambda\, Y_H^3(u), 
\end{equation}
as was shown in \cite{Bronnikov:2023lej}.
One of the solutions shown in Fig. \ref{Higgsfig} indicates that the Higgs field also exhibits sharp maxima on the branes.%

 {The Higgs distribution $H$ can be split as $H(x,y)=H_1(x,u)+H_2(x,u)$, see general expression \eqref{Psi12}. Recall that the distributions $H_1(x,u)$ and $H_2(x,u)$ act as independent fields as is discussed in \ref{spread} and \ref{floc}. The function $H_1(x,u)$ decreases rapidly far from the endpoint $u_1$ so that the observer on the brane-1 perceives the Higgs field as $H(x,y)=H_1(x,u)=h_1(x)Y^{(1)}_{H}(u)$, {see Fig.\ref{Higgsfig} left part of the plot}. The function $H_2(x,u)$ quickly decreases far from the endpoint $u_2$ and a possible observer on the brane-2  perceives the Higgs field as $H(x,y)=H_2(x,u)=h_2(x)Y^{(2)}_{H}(u)$. }

The expressions for the Higgs mass and the coupling constant as functionals of the function $Y_H(u)$ were obtained in \cite{Bronnikov:2023lej}. Applied to the brane-1, these parameter are as follows
\begin{eqnarray}\label{mh}
&&m^2_h ={v}_{\n-1} \int_{u_{1}}^{u_1+\Delta} \Bigl(-(\partial_u Y_H)^2 +\nu\, Y_H^2(u)\Bigr)\e^{4\gamma(u)}r^{\n-1}(u)\, du,
 \\         
&&\lambda_h = {v}_{\n-1} \int_{u_{1}}^{u_1+\Delta}
\lambda\,Y_H^4(u)\,\e^{4\gamma(u)}r^{\n-1}(u)\, du ,\quad 
{v}_{\n-1} \equiv \int d^{n-1} y \sqrt{|{g}_{n-1}|} =\dfrac{2\pi^{\tfrac{\n}{2}}}{\,\Gamma\left(\tfrac{\n}{2}\right)}.   \label{lh}
\end{eqnarray}
Here $\nu ,\, \Lambda$ are the initial physical parameters of the Higgs field in D-dim, ${v}_{\n-1}$ is the volume of the $(\n-1)$-dimensional sphere. 

 {The effective width of the brane, denoted by $\Delta$, is defined as the characteristic size of the extra dimension where the energy density of the localized fermion and the Higgs fields are non-negligible. The interaction between the Higgs field $H$ and the fermions $\Psi$ is described by vertex \eqref{SHe}.
Since the Higgs field and the fermion field $\Psi$ are tightly localized near the brane  $u_1$ (e.g., $u_1=0$), the main contribution to the integral over the extra dimension is restricted to the interval $(u_1, u_1 + \Delta)$.
Therefore, the 4D parameters of the Higgs field $m_h, \lambda_h$ are functionals of the distribution $Y^{(1)}_{H}(u)$ for the brane-1 observer, see Fig.\ref{Higgsfig}. The same is true for the Higgs vacuum average}
\begin{equation}\label{vh}
v_h=\frac{m_h}{\sqrt{2\lambda_h}}.
\end{equation}

 {These arguments can be applied to the brane-2 with the substitutions  $Y^{(1)}_{H}(u)\to Y^{(2)}_{H}(u)$ and the interval of integration $(u_1, u_1 + \Delta)\to (u_2-\Delta \to u_2)$ in expressions \eqref{mh} and \eqref{lh}. Hence, the vacuum average is different on both branes. 
}

It was declared from the beginning that the model is free from initially small physical parameter values. This means that the integrals as in \eqref{mh} and \eqref{lh} are of the order of unity if no special effort was made. The fine-tuning should be done in the expression \eqref{mh} to obtain the observable Higgs mass and its vacuum average.
As shown in the paper \cite{Bronnikov:2023lej}, the vacuum average can be adjusted to the known value $v_h\sim 10^{-17}m_{Planck}$ by an appropriate selection of the classical distribution $Y_H(u)$ over the extra dimensions. In the framework of the two-brane model discussed here, this procedure can be done for the brane-1 where the observer is assumed. This means that  $Y_H(u)$ should be substituted by $Y^{(1)}_{H}(u)$ in formulas \eqref{mh} and \eqref{lh}. The Higgs vacuum average on the brane-2 remains uncertain.

In the Standard Model, the fermion masses are proportional to the Higgs vacuum average $v_h$. The fermion-Higgs vertex in $D$ dimensions has the form
\begin{equation}\label{SHe}
    S_{He}=\int d^D X \sqrt{g_D}f_E \bar{\Psi}_LH \Psi_R=\int d^{D-1} X  \int du\sqrt{g_D}f_E \bar{\Psi}_LH \Psi_R
\end{equation}
where ${\Psi}_L$ is a left doublet of the SU(2) group, $\Psi_R$ is a right singlet and $H$ is the Higgs field. Below we show that the fermion masses and their coupling constant with the Higgs are different on both branes in 4D.

The decompositions \eqref{Psi12} in  the form \eqref{decomp}
\begin{eqnarray}
  &&  \Psi_L^{(1,2)}=\psi_L(x)Y^{(1,2)}_L(u), \\
  && \Psi_R^{(1,2)}=\psi_R(x)Y^{(1,2)}_R(u) 
\end{eqnarray}
of the fermions in D-dim on brane-1 (marked as $Y^{(1)}_{L,R}$) and brane-2 (marked as $Y^{(2)}_{L,R}$) leads to the following action
  \begin{equation}
  S^{(1,2)}_{He}=\int d^4 x \sqrt{g_4}f_e^{(1,2)} \bar{\psi}_L(x) h(x) \psi_R(x) 
  \end{equation}
after their substitution in \eqref{SHe}. All functions acting in the 4D space are assumed to be normalized to obtain the standard form of the kinetic terms.
The value of the 4D coupling constant
\begin{equation}\label{f12}
 f^{(1,2)}_e = v_{n-1}f_E\int due^{4\gamma(u)}r(u)^{n-1} Y^{(1,2)}_L(u)Y^{(1,2)}_R(u)Y^{(1,2)}_{H}(u)
\end{equation}
depends on which brane the fermion is located because the fermion distributions reveal a sharp peaks: $Y^{(1)}_{L,R}(u)$ on the brane-1 and $Y^{(2)}_{L,R}(u)$ on the brane-2, see Fig.\ref{fermifig}. The metric functions $\gamma(u), r(u)$ and the distribution $Y_H(u)$ are unique near each brane, so the coupling constant values  $f^{(1,2)}_e $ are different there.

If we live on brane-1, the SM in 4 dimensions and the parameter $f^{(1)}_e$  must have the observable value  $f^{(1)}_e =m_e/v^{(1)}_h\sim 10^{-6}$, see \cite{Workman:2022ynf}. This value is very small and hence needs to be fine-tuned in addition to the vacuum average $v_h$. This can be done by selecting appropriate functions $Y^{(1)}_{L,R}(u)$ in \eqref{f12}. 
    
In this research, we concentrate on the fields located on the brane-2, where the vacuum average $v^{(2)}_h$, the coupling constant $f^{(2)}_e $ and hence, the electron mass $m_2\sim f^{(2)}_e v^{(2)}_h$ remain uncertain varying in a wide range due to the absence of the fine-tuning. 

\subsection{Stability}

 {The study of static solutions stability in extra dimensional models based on nonlinear $f(R)$ gravity is extremely complicated problem. The results obtained up to now are based on specific assumption on the analytical form of solutions.
}

 {Tensor-mode stability for the brane background is analyzed in \cite{Xu:2014jda, Xu:2022xxd, Liu:2009dt}. The stability of the compact static metric is usually connected to the radion field \cite{Koyama:2002nw} which is defined as the physical distance between two branes which  leading to interesting results if the metric is maximally symmetric or in the models inspired by the Rassel-Sundrum one.  In the model developed here, the situation is more complicated because of the absence of appropriate metric symmetries and the numerical character of solutions interpreted as the thick branes. Our numerical simulations made in \cite{Bronnikov:2020tdo} indicate that perturbations are attenuated with time.  }

 {The sensitivity of the model to additional conditions is also worth noting. In this approach, the brane structure emerges from solutions to the field equations subject to specific additional conditions---namely, the metric functions and their derivatives are fixed at a particular coordinate $ u $. The nontrivial fact is that vast majority of solutions terminate at certain coordinates, exhibiting the properties of static branes. A slight variation in these additional conditions leads to a correspondingly small variation in the metric functions, while the presence of the branes themselves persists. Massive, radion-like perturbations appear only at extremely high energy scales, which lie far beyond the scope of the present study. }
	
 {The analogy with the four-dimensional collapse of dust to form a black hole provides strong intuitive support for our model. Indeed, the final mass of a black hole is determined by the initial distribution of dust that collapses to create it. A small variation in the initial dust distribution leads to a correspondingly small variation in the black hole mass, in direct analogy to the behavior of the extra-dimensional metric functions discussed above. 
	Evidently, the resulting black hole masses form a continuous, parameterized set of static solutions. The same can be said of the extra-dimensional metric functions. We also note that black holes themselves represent metastable states due to Hawking radiation.}

\section{Matter fields in the uninhabited brane. \newline Problems and prospects.}

In this section, we investigate potential observational signatures of our extra-dimensional model proposed earlier. As established above, the energy scale hierarchy problem represents a fundamental theoretical challenge that appears resolvable only within brane-1. This framework naturally predicts that brane-2 hosts superheavy partners to all Standard Model particles, with masses sufficiently large to: (i) serve as viable dark matter candidates, and/or (ii) potentially produce other distinctive observational signatures. 

This scenario could also lead to fundamentally different astrophysical evolution in each brane: while complex structures like stars might be prohibited in brane-2, stable hydrogen-like matter could still form if QCD dynamics operate similarly across both branes.


\subsection{Dark matter components?}\label{dark}
One of the main goals of the cosmology is to understand the nature of the dark matter. Special efforts have been applied to explain the properties of this phenomenon and the origin of dark matter \cite{2006PhRvD..73a5001B,2007PhR...453...29H,2006PZETF...83..3,Arbuzova:2021etq,Belotsky:2014kca,2003PhRvD..68j3003B}. 
Many theoretical models as well as the observational evidences can be found in the reviews \cite{Beylin:2019gtw,Arbey:2021gdg,Misiaszek:2023sxe}. 
The idea of extra dimensions is also attracted in the framework of this discussion, \cite{Chu:2011be,2005JCAP...07..001B,2009MPLA...24..667K,2006PhRvD..73a5001B}. Some of them are treated the brane idea  \cite{Boehmer:2007xh,SHAHIDI_2011} which is natural basis for production of weakly interacted particles. 

In this paper, we have demonstrated that brane-2 contains heavy partners of all Standard Model (SM) particles and their corresponding antiparticles. Among these, heavy neutrinos stand out as promising dark matter candidates. A detailed discussion on this topic can be found in \cite{Abdallah:2024npf,Barman:2022scg}.

A more thorough discussion is necessary to address the implications of charged particles, such as electrons in the dark matter context. At first glance, it may seem that the dark matter cannot contain charged particles due to their significant interaction with photons (see also the discussion in \cite{Kamada2016EffectsOE}). This holds true for observable particles like electrons. However, the situation changes if the particles are sufficiently massive.
Indeed, let us estimate the Compton effect on brane-2, i.e. the photon scattering on the non-relativistic heavy electron. This cross section depends on a particle mass $m$ as
\begin{equation}\label{se}
    \sigma_e\sim \frac{1}{m^2}\sim 10^{-26} \text{cm}^2.
\end{equation}
The second approximate equality is the known experimental result for the standard electron mass $m_e$.
The same cross section for the heavy electrons is smaller by the factor $(m_e/m_2)^2$ than the standard one. Here $m_e$ and $m_2$  are the masses of the ''normal'' electron and the ''heavy'' electrons.
We can estimate the mass of the heavy electrons supposing that their Compton cross section should be smaller than the lower observational bound $\sim 10^{-44}$ cm$^2$ for the dark matter cross section,
\begin{equation}\label{comp}
   \sigma_2\sim \frac{1}{m_2^2}\sim \frac{m_e^2}{m_2^2} \sigma_e\lesssim 10^{-44}\text{cm}^2 . 
\end{equation}
It is easy to see that the heavy electron mass should be really large, 
\begin{equation}\label{m2limit}
    m_2 > 10^{9}m_e\sim 10^6 \text{GeV},
\end{equation}
where the value of $\sigma_e$ is taken from \eqref{se}.
The mass estimate \eqref{m2limit} for heavy electrons on brane-2 allows one to consider them as the dark matter candidate.

Nevertheless, caution is warranted. Scattering between ordinary electrons and their massive counterparts is unsuppressed, as evidenced by the effective vertices in \eqref{fe}. This interaction could potentially distort the CMB and lead to unacceptable accumulation of the dark matter in the form of heavy electrons in galactic centers.

Charged particles, being extremely massive could also form bonded states like the positronium and the hydrogen. The positronium life-time $\sim m^{-1}$, ($m$ is the heavy electron mass on the brane-2) and hence decay instantly after formation. The concentration of the heavy electrons is significantly suppressed, yet it remains nonzero. 
Heavy hydrogen cannot be a viable dark matter candidate if strong interaction remains unchanged in brane-2.

 {The properties of particles in the uninhabited brane are completely unknown. While Standard Model particles exist there, their masses, coupling constants, and concentrations are undetermined.
These unknown brane-2 particles form specific subdominant components of dark matter and might also explain certain observed phenomena, which could, in turn, help clarify their properties keeping in mind restrictions imposed by known and firmly established constraints from CMB photon scattering, for example. 
Below we perform very preliminary discussion on one of such phenomena.}

\subsection{Ultra-high-energy cosmic rays}
Ultra-high-energy cosmic rays (UHECR) of the energy $10^{11}$GeV \cite{ParticleDataGroup:2024cfk} are one of the most intriguing mysteries of modern astrophysics. 
So far, only several protons of such energy were definitely registered. Despite numerous attempts to explain this phenomenon, its origin remains largely unknown. The discussion on the origin of the high-energy photons of PeV energy can be found in \cite{2011APh....35...28B}.

The two-brane model discussed here provides us with a possible explanation of this phenomenon based on the heavy electron-positron annihilation into a pair of quarks. The part of action \eqref{fe} contains three vertexes. The first term contains two fermion settled on brane-1 interacting with an electromagnetic field. The second one contains two fermions settled on the neighboring brane-2 coupled with the same electromagnetic field. So, the process describing the annihilation fermions on brane-2 into photon with the following creation of two other fermions on the brane-1 can be realized.
The last term in \eqref{fe} is proportional to the extremely small coupling constant $\alpha_{12}$ and is hardly significant.
The energy $\varepsilon$ of the emitted quarks on the brane-1 is of the order of the heavy electron mass $m_2$
\begin{equation}
    \varepsilon \sim m_2
\end{equation}
provided that they are non-relativistic. This allows us to estimate the mass of the electrons settled on the brane-2 if we suppose that UHECR are the result of the heavy electron-positron annihilation,
\begin{equation}\label{m2}
    m_2\sim 10^{11}\text{GeV}.
\end{equation}
This value is consistent with the lower boundary limits \eqref{m2limit}. Given the extremely small cross-section, such events occur rarely, as outlined earlier. For a rigorous analysis of superheavy charged particles in this framework, see \cite{Arbuzova:2024uwi} 

A similar mechanism used to explain high-energy photons encounters difficulties. The highest observed photon energy is about $10^6$ GeV, and the mass of a heavy electron must be of this order of the magnitude. However, this value lies below the threshold defined in \eqref{m2limit}, so some efforts should be applied to connect this effect to the particles located in the uninhabited brane.

\section{Conclusion}
Two-brane extra dimensions play an important role in theoretical research starting from the Randall-Sundrum model.

 In this research, we study the distribution of fields using the two-brane model developed in \cite{Popov:2024nax}. This model has revealed its ability to explain the electroweak hierarchy problem, the restoration of the Starobinsky inflationary model \cite{Starobinsky:1980te}, and the smallness of the cosmological constant \cite{Bronnikov:2023lej}.


Here we focus on the fields distribution over the extra dimensions. The extra-dimensional regions where fermions are localized are referred to as (thick) branes. It is shown that fermions split into two independent effective fields, localized on adjacent branes. This bifurcation process accompanies the evolution of the Universe from energy levels above the inflationary scale to the present-day state. 
Therefore, the known Standard model particles are presented in both branes.
The conditions under which the fermions and the Higgs field are independent at neighboring branes are attentively discussed in the subsection \ref{spread}. 
At the same time, the gauge fields are uniformly distributed in the extra dimensions, preserving the charge universality in 4D physics.

The emergence of complex structures requires a hierarchy of energy scales and associated fine-tuning \cite{2007unmu.book..231D}. In a brane-world model, the delicate process of parameter selection for the fine-tuning can be accomplished on a single brane, the one inhabited by the observers (brane-1).  This means that the masses of the particles, partners of Standard Model particles in the other brane should have ''natural'' values, i.e. be of the order of the energy scale at which our Universe was formed, i.e. extremely massive. The Higgs mechanism of mass creation on both branes is discussed in subsection \ref{Higgs}. In addition, a secondary brane (brane-2) are lack the fine-tuning, precluding the formation of stellar-like structures.

The stable matter located on the brane-2 consists of the charged particles such as 3 generations of leptons and the protons as well as neutral particles like the neutrinos and the atoms of hydrogen.
Neutral particles are obvious candidates for the role of dark matter. The extremely high mass of heavy charged leptons on brane-2 is the key factor for the significant reduction of the Compton scattering cross section. This allows the heavy charged leptons to be considered as a component of dark matter, albeit with a likely small contribution due to stringent observational constraints.
 
It is also shown that the fermions located on the different branes can exchange via photons and gravitons and thus be a source of ultra-high energy particles.

The model of extra dimensions discussed here seems to be a promising framework that deserves further development. It could also serve as a basis for explaining phenomena such as the positron anomaly \cite{Belotsky:2018vyt} and the smallness of the neutrino mass.

\section*{Acknowledgements}

The author is grateful to K. Belotsky and R. Konoplich for fruitful discussions.
The work was partly funded by the Ministry of Science and Higher Education of the Russian Federation, Project "Studying physical phenomena in the micro- and macro-world to develop future technologies" FSWU-2026-0010.
The research presented in Sect.3,4 was also partially carried out in accordance
with the Kazan Federal University Strategic Academic Leadership Program.

 {\section*{Conflicts of interests}
	The author confirm that there are no any potential conflict(s) of interest like employment, consulting fees, research contracts, stock ownership, patent licenses, honoraria, advisory affiliations etc. 
}

\printbibliography[title={References}, heading=bibintoc]


\end{document}